\documentclass[aps,twocolumn,showpacs,preprintnumbers]{revtex4}

\usepackage{graphicx}  
\usepackage{subfigure}
\usepackage{multirow}

\usepackage{fancyhdr}
\usepackage{longtable}
\usepackage{parskip}
\usepackage[T1]{fontenc}
\usepackage{dcolumn}   

\usepackage{bm}        
\usepackage{amsfonts}  
\usepackage{amsmath}   
\usepackage{amssymb}   

\newcommand{\pwisein}{\left\{ \begin{array}{ll}}
\newcommand{\pwiseout}{\end{array}\right.}

\begin{document}

\title{\large\bf Two Quantum Proxy Blind Signature Schemes Based on Controlled Quantum Teleportation}
\author{ Qiming Luo$^1$, Tinggui Zhang$^{1, 2,\dag}$, Xiaofen Huang$^{1,2}$ and Naihuan Jing$^{3,4}$}
\affiliation{ ${1}$ School of Mathematics and Statistics, Hainan Normal University, Haikou, 571158, China \\
${2}$ Key Laboratory of Data Science and Smart Education, Ministry of Education, Hainan Normal University, Haikou, 571158, China \\
$3$ College of Sciences, Shanghai University, Shanghai, 200444, China \\
$4$ Department of Mathematics, North Carolina State University, Raleigh, NC 27695, USA\\
$^{\dag}$ Correspondence to tinggui333@163.com\\}
\date{}
\bigskip

\begin{abstract}
We present a scheme for teleporting an unknown, two-particle
entangled state with a message from a sender (Alice) to a receiver
(Bob) via a six-particle entangled channel. We also present another
scheme for teleporting an unknown one-particle entangled state with
a message transmitted in a two-way form between the same sender and receiver via a five-qubit cluster state. One-way hash functions,
Bell-state measurements, and unitary operations are adopted in these
two schemes. Our schemes use the physical characteristics of quantum
mechanics to implement delegation, signature, and verification processes.
Moreover, a quantum key distribution protocol and a one-time pad are
adopted in these schemes.

\end{abstract}

 \maketitle

\section{Introduction}
Following the in-depth development of quantum key distribution (QKD)
\cite{benn,accc,addd}, various quantum cryptographic
protocols have been developed vigorously, such as the quantum secure
direct communication \cite{lgll,aeee,afff,aggg}, quantum private
query \cite{ahhh,aiii,ajjj}, quantum secret
sharing \cite{akkk,alll,ammm}, quantum multiparty secure
computations \cite{annn,aooo}, quantum
authentication \cite{appp,aqqq,arrr}, quantum signature \cite{attt},
and others. To ensure the safety of quantum cryptographic
protocols, some attacking strategies on quantum cryptography protocols
have been proposed, such as intercepted-resend \cite{auuu,avvv}, entanglement-swapping \cite{awww,axxx}, teleportation \cite{ayyy},
dense-coding \cite{azzz,bccc,bddd}, channel-loss \cite{beee}, denial-of-service \cite{bfff,bggg},
correlation-extractability \cite{bhhh,biii}, trojan-horse \cite{bjjj,bkkk}, information-leakage \cite{blll},
collusive attacks\cite{bmmm}, and others. These effective attack strategies
helped the design of new quantum cryptography schemes that are more
secured.

In recent years, many different quantum signature schemes have been
proposed for different application environments, including (but not limited to) quantum blind \cite{bnnn,booo}, quantum proxy
\cite{bppp,bqqq}, quantum multiple \cite{brrr,bsss}, quantum group \cite{bttt,buuu}, and others. In 2013, Zhang et al. proposed a secure quantum group signature
scheme based on the Bell state \cite{bvvv}, and they also presented the cryptanalysis of the quantum group signature
protocol \cite{bwww}. These signature schemes can be applied to
e-commerce or other related fields.

The digital signature, which was independently introduced by
Diffe \cite{aaaa} and Merkle \cite{bbbb}, has been an important branch
of cryptography. Digital signatures constitutes a very important research
topic in classic cryptography as it has many applications in our
real life. The security of classical signature schemes is based on
complex mathematical problems, such as the factorization and discrete
logarithm problems. However, classic signatures are becoming increasingly vulnerable as quantum algorithms develop further.
Fortunately, in quantum information processing and computation,
quantum cryptography can provide secure communication based on
quantum mechanics, especially when based on the no-cloning theorem. Inspired by these properties, some progress has been achieved in the field of quantum signatures \cite{cccc,dddd,eeee}.

Proxy signatures, which allow the original signer to authorize the proxy signer to sign messages on his behalf, plays a key role
in cryptography. Since Mambo et al. \cite{ffff} proposed the concept of proxy signature in 1996, some signature schemes have been proposed. Wang et al. \cite{gggg} proposed a
one-time proxy signature scheme without decoherence. Yang et al. \cite{hhhh} subsequently indicated that this scheme could not meet the security requirements of non-forgery and
nonrepudiation. However, Wang et al. \cite{iiii} showed in 2015 that this conclusion was not appropriate. Recently, Cao et al.
\cite{jjjj,kkkk} presented two quantum proxy signature schemes based
on some genuine six- and five-qubit entangled states. However, Zhang et al. \cite{llll} suggested that receivers could forge valid signatures. In 2015, Tian et al. \cite{mmmm}
presented a quantum multi-proxy blind signature scheme based on four-qubit entangled states with fewer resources.

The blind signature is a special digital signature that can protect the anonymity of the message owner to ensure privacy. In blind signatures, the message owner can always obtain
 the real signature of his message, even if the signer knows nothing about the content of his signature. Blind signatures can be classified into weak and strong blind signatures based
  on whether the signer can track the message owner. Chaum proposed the first blind signature scheme in 1983 based on the complexity of factoring large integers \cite{nnnn}. However,
it could be easily broken with the emergence of quantum computers.
In 2010, Su et al. \cite{oooo} presented a weak blind quantum
signature scheme based on a two-state vector formalism. They used EPR pairs to implement the signature process, and applied the characteristics of the two-state vector form to implement the
verification scheme. In recent years, quantum signature schemes have received extensive attention.

In this study, we propose two quantum proxy blind signature
schemes based on controlled quantum teleportation. The first scheme
takes advantage of the correlation of the maximal EPR and
four-qubit entangled states, Bell-state
measurements (BMs), $\{|0\rangle,|1\rangle\}$ basis measurements,
Hadamard operation, quantum-key distribution, and unitary operations.
In this scheme, two particles are delivered to the original signer
Alice. The proxy signers Charlie and David hold one particle each. The remaining two particles are distributed to the
verifier Bob. The second scheme is bidirectional, taking advantage of the
correlation of the five-qubit cluster state, BMs, $\{|+\rangle,|-\rangle\}$ basis measurements,
quantum-key distribution, and unitary operations. In this scheme, two
particles are delivered to the original signer and verifier Alice.
The proxy signer Charlie holds one particle. The other two particles
are distributed to the verifier and original signer Bob. We used the
quantum key distribution and a one-time pad to guarantee the
unconditional security and signature anonymity. It is shown to be
unconditionally secured, i.e., may not be forged or modified in any
way by the receiver or attacker. In addition, it may neither be
disavowed by the signatory, nor denied by the receiver.

\section{Controlled Quantum Teleportation of the First Scheme}
Our multiproxy blind signature scheme is based on the controlled quantum teleportation that uses the maximal EPR state and maximal four-qubit entangled states as its quantum channel. It is given by
\begin{eqnarray}   \label{eq}
|\psi\rangle_{3456}&=&{\frac{1}{2\sqrt{2}}}(|0000\rangle-|0111\rangle+|1001\rangle-|1110\rangle \nonumber \\
&+&|0110\rangle+|0001\rangle+|1111\rangle+|1000\rangle)_{3456},  \\
|\psi\rangle_{78}&=&{\frac{1}{\sqrt{2}}}(|00\rangle+|11\rangle)_{78}.
\end{eqnarray}
Suppose that the quantum state of particle carrying message in Alice is
\begin{eqnarray}   \label{eq}
|\phi\rangle_{12}=(\alpha|00\rangle+\beta|01\rangle+\gamma|10\rangle+\delta|11\rangle)_{12},
\end{eqnarray}
where the unknown coefficients $\alpha$, $\beta$, $\gamma$ and
$\delta$ satisfy $|\alpha|^2+|\beta|^2+|\gamma|^2+|\delta|^2=1$.

This controlled quantum teleportation scheme involves the following four
partners: the sender (Alice), two controllers (Charlie and David), and the
receiver (Bob). Alice holds particles (1,2,3,7), Bob holds particles
(6,8), and Charlie and David own particles 4 and 5. Therefore, the
quantum state $|\Psi\rangle$ of the entire system composed of
particles (1,2,3,4,5,6,7,8) is the following
\begin{eqnarray}   \label{eq}
|\Psi\rangle=|\phi\rangle_{12}\otimes|\psi\rangle_{3456}\otimes|\psi\rangle_{78}.
\end{eqnarray}
The steps of the controlled quantum teleportation are as follows (the detailed process is described in reference \cite{pppp}).

(1) Alice performs BMs on particles (1,3) and (2,7). These are 16 possible outcomes.

(2) Alice sends her measurement outcomes to Bob through a secure
quantum channel. Bob then performs a corresponding unitary operation
on particles (6,8).

(3) If Charlie and David agree with Alice and Bob to complete their
teleportation process, they perform the Hadamard operation on their particles, respectively. Subsequently, they perform a \{$|0\rangle$,$|1\rangle$\} basis
measurement on their particles. They then send their
measurement outcomes to Bob through a secure quantum channel.

(4) When the two measurements are the same, Bob will obtain the quantum
state transferred from Alice. When the two measurements are
different, Bob just needs to apply the unitary operation $\sigma_x$
on particle 6, whereupon the same receiver obtains the quantum state transferred from Alice.

\section{Quantum Multiproxy Blind Signature Scheme}
In our scheme, the participants are defined as follows.

(i) Alice: the original signer, (ii) Bob: the message receiver, (iii) Charlie and David: two proxy signers, and (iiii) Trent: the message verifier and trusted party.

The detailed procedure of our scheme can be described as follows.
\subsection{Initial Phase}
(i) QKD. Alice shares the secret key $K_{AB}$ with
Bob. In addition, Bob establishes the secret keys $K_{BC}$ with Charlie and
$K_{BD}$ with David. Moreover, Trent shares the secret keys $K_{TA}$ with
Alice, $K_{TB}$ with Bob, $K_{TC}$ with Charlie, and $K_{TD}$ with
David. These distribution tasks can be fulfilled via QKD protocols,
which have been proved unconditionally safe.

(ii) Quantum Channel Establishment. Bob produces ${t+l}$ quantum
states $|\psi\rangle_{3456}$ and $|\psi\rangle_{78}$. He sends
particles (3,7) to Alice, particle 4 to Charlie, and particle 5 to
David, leaving particles (6,8) to himself.

(iii) To ensure security of the quantum channel, Bob
arranges eavesdropping checks.

\subsection{Blinding the Message Phase}
(i) Alice converts her message $m$ into an N-bit sequence and records $m=\{m(1),m(2),...m(j),\dots,m(N)\}$. Subsequently, Alice sends the binary sequence $m$ to Trent.

(ii) Alice transforms $m$ to $H(m)$ by using a Hash function, where $H$:
$\{0,1\}^N \to \{0,1\}^n(N \gg n)$, and Trent also knows the Hash
function $H$. She then blinds the massage based on $M(i)=K_{TA}\oplus
m(i), i=1,2,\dots,n$, where $\oplus$ is the XOR operation. She can
identify an appropriate Hash function, such that $n$ is an even umber. Let
$l={\frac{n}{2}}$.

(iii) Alice produces ${\frac{n}{2}}$ quantum states $|\phi\rangle_{12}=(\alpha|00\rangle+\beta|01\rangle+\gamma|10\rangle+\delta|11\rangle)_{12}$,
denoted as $|\phi\rangle_{12}(1)$, $|\phi\rangle_{12}(2)$, \dots, $|\phi\rangle_{12}({\frac{n}{2}})$. If $m(i)=00, |\phi\rangle_{12}=|00\rangle$;
If $m(i)=01, |\phi\rangle_{12}=|01\rangle$; if $m(i)=10, |\phi\rangle_{12}=|10\rangle$; if $m(i)=11, |\phi\rangle_{12}=|11\rangle$.

\subsection{Authorizing and Signing Phases}
In our scheme, we used a one-time pad as the encryption algorithm to
ensure the unconditional security.

(i) If Alice agrees with Charlie and David as her proxy signers to sign
the message, she will help to perform controlled teleportation.
Alice performs the BMs on particles (1,3) and
(2,7) and records the measurement outcomes as $S_A$.
She then encrypts $S_A$ with the key $K_{AB}$ to obtain the secret message
$E_{K_{AB}}\{S_A\}$. Alice sends the message $E_{K_{TA}}\{S_A\}$ to Trent via the quantum channel. Similarly, Alice sends the
message $E_{K_{AB}}\{S_A\}$ to Bob via the quantum channel.

(ii) After Bob has received the message $E_{K_{AB}}\{S_A\}$, he
decrypts it with his $K_{AB}$ to get the message $S_A$. Bob then
performs corresponding unitary operations on particles (6,8). After
this, he encrypts $S_A$ with the keys $K_{BC}$ and $K_{BD}$ to obtain the secret messages $E_{K_{BC}}\{S_A\}$ and $E_{K_{BD}}\{S_A\}$, respectively. Bob sends the
 messages $E_{K_{BC}}\{S_A\}$ and $E_{K_{BD}}\{S_A\}$ to Charlie and David, respectively, via the quantum channel.

(iii) After Charlie and David have received the messages
$E_{K_{BC}}\{S_A\}$ and $E_{K_{BD}}\{S_A\}$, they decrypt it with
their keys $K_{BC}$ and $K_{BD}$ to obtain the message $S_A$. They then
perform the Hadamard operation on their particles. Subsequently,
they perform a \{$|0\rangle$,$|1\rangle$\} basis measurement on
their particles, and they note the measurement outcomes
as $S_C=\{c(1),c(2),...c(i),\dots,c({\frac{n}{2}})\}$ and
$S_D=\{d(1),d(2),...d(i),\dots,d({\frac{n}{2}})\}$. Charlie then encrypts \{$S_A$, $S_C$\} with the use of the key $K_{BC}$ to obtain the secret
message $S_1=E_{K_{BC}}\{S_A, S_C\}$, and David also encrypts
\{$S_A$, $S_D$\} with the key $K_{BD}$ to obtain the secret message
$S_2=E_{K_{BD}}\{S_A, S_D\}$. They send the messages $S_1$ and $S_2$
to Bob as the proxy authorization. Similarly, they send \{$S_A$,
$S_C$\} and \{$S_A$, $S_D$\} to Trent by $K_{TC}$ and $K_{TD}$.

\subsection{Verifying Phases}
(i) Bob receives the messages $S_A$, $S_C$ and $S_D$ by using the keys
$K_{BC}$ and $K_{BD}$ to decrypt $S_1$ and $S_2$. If $S_A$ in the
proxy signature does not match that sent by Alice, the
signature verification process will be terminated, and the signature
will be declared invalid. Otherwise, the process continues based on the following steps.

(ii) According to $S_C$ and $S_D$, Bob performs an appropriate
unitary operation on particle 6 to replicate the unknown state
$|\phi\rangle_{12}$ which carries the messages.
 He then sends the state $|\phi\rangle_{12}$ to Trent.

(iii) Trent encodes the quantum state $|\phi\rangle_{1'2'}$  to
obtain the message $M'$. If $M'=M$, he confirms a series of
signatures and messages $(m, S_C, S_D)$. Otherwise, Trent rejects
it.

\section{Security Analysis and Discussion of the First Scheme}
In this section, we will demonstrate that our scheme meets the following security requirements in accordance with reference \cite{llll}.
\subsection{Message Blindness}
Alice blinded the message $m$ to $M$. Proxy signers Charlie and David could not obtain the contents of the messages because Alice used hash functions and XOR operations to blind them.
Therefore, Charlie and David could not know the content of the information she signed.
\subsection{Impossibility of Denial}
In this scheme, we prove that Alice cannot refuse her delegation, and Charlie and David cannot refuse their signatures. If the signature is verified, the signer cannot reject its signature or
message for the following reasons: 1) the signature is
encrypted by the key, and 2) the entangled states have stable coherence among particles. Only by measuring their own particles they can achieve teleportation. All keys are distributed based on
 the QKD protocol, which has been unconditionally proved. All messages are sent through secure quantum channels. Therefore, Alice cannot refuse her delegation, nor can Charlie and David refuse their signatures.
\subsection{Impossibility of Forgery}
Firstly, we assume that Trent is completely credible. We can prove that Charlie and David cannot forge Alice's signature. In our scheme, trusted Trent can prevent these attacks. Suppose Charlie tries to
forge Alice's signature. Even if he knows the key, he cannot forge any signature because Charlie's forgery attack will be detected by Trent. Similarly, David could not forge Alice as
Charlie's signature. In other words, internal attackers cannot forge their signatures.

Secondly, we assumed that there was an external attacker (Eve). Eve cannot obtain the secret key. Therefore, she cannot forge the signatures of Alice, Charlie, and David. We assumed that Eve could obtain the secret
key randomly and generate a valid signature with the
probability ${\frac{1}{2^n}}$ tending to zero when $n$ tends to infinity.
At the same time, we realized the eavesdropping detection function in the construction stage of the quantum channel. Therefore, the signing process will only continue when the channel is secure; otherwise, the signing process will terminate. Therefore, Eve cannot forge their signatures.

\section{Controlled Quantum Teleportation of the Second Scheme}
Our two-way proxy blind signature scheme is based on controlled quantum teleportation that uses the five-qubit cluster state as its quantum channel. It is given by,
\begin{eqnarray}   \label{eq}
|\phi\rangle_{12345}={\frac{1}{2}}(|00000\rangle+|00111\rangle+|11101\rangle+|11010\rangle).
\end{eqnarray}
Suppose that the quantum state of particle carrying message in Alice is
\begin{eqnarray}   \label{eq}
|\phi\rangle_{A}&=&(\alpha|0\rangle+\beta|1\rangle)_{A},  \nonumber  \\
|\phi\rangle_{B}&=&(\gamma|0\rangle+\delta|1\rangle)_{B},
\end{eqnarray}
where the unknown coefficients $\alpha$, $\beta$, $\gamma$ and
$\delta$ satisfy $|\alpha|^2+|\beta|^2=1$ and
$|\gamma|^2+|\delta|^2=1$.

This controlled quantum teleportation involves the following three partners: the sender and receiver Alice, sender and receiver Bob, and the controller Charlie. Alice held particles (1,5,A),
 Bob held particles (2,3,B), and Charlie owned particle 4. Thus, the quantum states $|\Psi\rangle$ of the entire system takes the following form
\begin{eqnarray}   \label{eq}
|\Psi\rangle=|\phi\rangle_{B}\otimes|\phi\rangle_{A}\otimes|\phi\rangle_{12345}.
\end{eqnarray}
The steps of the controlled quantum teleportation are as follows (the detailed process is described in reference \cite{qqqq}).

(1) Alice performs BMs on particles (A,1) and
Bob performs BMs on particles (B,3) respectively.
These are 16 types of possible outcomes.

(2) Alice sends her measurement outcomes to Bob and Charlie through
the secured quantum channel. Bob sends his measurement outcomes to Alice and Charlie.

(3) If Charlie agrees with Alice and Bob to complete their two-way
teleportation processes, he performs a \{$|+\rangle$,$|-\rangle$\} basis
measurement on particle 4. He then sends his measurement outcomes to
Alice and Bob through the secured quantum channel.

(4) Bob performs a corresponding unitary operation on particle 2,
whereupon Bob receives the quantum state transferred from Alice. The same applies for Alice who receives the quantum state on particle 5 transferred from Bob.

\section{Quantum Two-way Proxy Blind Signature Scheme}
In our scheme, participants are defined as follows:

(i) Alice and Bob: the original signers, (ii) Charlie: the proxy signer, (iii) Trent: the message verifier and the trusted party.

The detailed procedure of our scheme can be described as follows.
\subsection{Initial Phase}
(i) QKD. Alice shares the secret key $K_{AB}$
with Bob and shares the secret key $K_{AC}$ with Charlie. In
addition, Bob establishes the secret key $K_{BC}$ with Charlie.
Moreover, Trent shares the secret keys $K_{TA}$ with Alice, $K_{TB}$
with Bob, $K_{TC}$ with Charlie. These distribution tasks can be
fulfilled via QKD protocols, which have been proven to be unconditionally
secured.

(ii) Quantum Channel Establishment. Bob produces ${p+q}$ quantum
states $|\phi\rangle_{12345}$. He sends particles (1,5) to Alice,
particle 4 to Charlie, and leaves particles (2,3) to himself.

(iii) To ensure the security of the quantum channel, Bob
arranges eavesdropping checks.

\subsection{Blinding the Message Phase}
(i) Alice converts the message $m'$ into a K-bit sequence and records $m'=\{m'(1),m'(2),...m'(j),\dots,m'(K)\}$. Subsequently, Alice sends the binary sequence $m'$ to Trent.
Similarly, Bob converts the message $m''$ into an R-bit sequence and records $m''=\{m''(1),m''(2),...m''(j),\dots,m''(R)\}$. Subsequently, Bob sends the binary sequence $m''$ to Trent.

(ii) Alice converts $m'$ to $H'(m')$ by using the Hash function, where $H'$,
$\{0,1\}^K \to \{0,1\}^q(K \gg q)$, and Trent also knows the Hash
function $H'$. She then blinds the massage by $M'(i)=K_{TA}\oplus
m'(i), i=1,2,\dots,q$, where $\oplus$ is the XOR operation. Similarly, Bob converts $m''$ to $H'(m'')$ by using the
Hash function, where $H'$: $\{0,1\}^K \to \{0,1\}^q(K \gg q)$. He then blinds the massage by $M''(i)=K_{TB}\oplus m''(i),
i=1,2,\dots,q$.

(iii) Alice produces $q$ quantum states
$|\phi\rangle_{A}=(\alpha|0\rangle+\beta|1\rangle)_{A}$, denoted as
$|\phi\rangle_{A}(1)$,$|\phi\rangle_{A}(2)$,\dots,$|\phi\rangle_{A}(q)$. Similarly, Bob produces $q$ quantum states
$|\phi\rangle_{B}=(\gamma|0\rangle+\delta|1\rangle)_{B}$, denoted as
$|\phi\rangle_{B}(1)$,$|\phi\rangle_{B}(2)$,\dots,$|\phi\rangle_{B}(q)$.

\subsection{Authorizing and Signing Phase}
In our scheme, we used the one-time pad as the encryption algorithm to
ensure the unconditional security.

(i) If Alice agrees that Charlie can act as her proxy signer to sign the message, she will help the execution of controlled teleportation. Alice performs the BM on particles (A,1) and records the
measured outcomes as $S_A$. She then encrypts $S_A$ with the keys
$K_{AB}$ and $K_{AC}$ to obtain the secret messages $E_{K_{AB}}\{S_A\}$
and $E_{K_{AC}}\{S_A\}$. Alice sends the messages
$E_{K_{AB}}\{S_A\}$ to Bob and $E_{K_{AC}}\{S_A\}$ to Charlie via the quantum channel. Similarly, If Bob agrees that Charlie can act as his proxy signer to sign the message, he will help the execution of controlled
teleportation. Bob performs the BM on particles
(B,3) and records the measuring outcomes as $S_B$. He then encrypts
$S_B$ with the keys $K_{AB}$ and $K_{BC}$ to obtain the secret messages
$E_{K_{AB}}\{S_B\}$ and $E_{K_{BC}}\{S_B\}$. Bob sends the messages
$E_{K_{AB}}\{S_B\}$ to Alice and $E_{K_{BC}}\{S_B\}$ to Charlie via
the quantum channel.

(ii) After Charlie has received the messages $E_{K_{AC}}\{S_A\}$ and
$E_{K_{BC}}\{S_B\}$, he decrypts them with their keys $K_{AC}$ and
$K_{BC}$ to obtain the messages $S_A$ and $S_B$. Charlie then performs
a \{$|+\rangle$,$|-\rangle$\} basis measurement on particle 4, and
he notes the measured outcome as
$S_C=\{c(1),c(2),...c(i),\dots,c(q)\}$. Charlie encrypts
\{$S_A$, $S_C$\} with the use of the key $K_{BC}$ to obtain the secret message $S_1=E_{K_{BC}}\{S_A, S_C\}$, and also encrypts \{$S_B$, $S_C$\} with the use of the
 key $K_{BC}$ to obtain the secret message $S_2=E_{K_{BC}}\{S_B,S_C\}$. He then sends the messages $S_1$ to Bob and $S_2$ to Alice as the proxy authorization. Similarly,
  Charlie sends \{$S_A$, $S_C$\} and \{$S_B$, $S_C$\} to Trent by using $K_{TA}$ and $K_{TB}$, respectively.

\subsection{Verifying Phases}
(i) Bob receives the messages $S_A$, $S_C$ by using the key $K_{BC}$ to
decrypt $S_1$. If $S_A$ in the proxy signature does not match that sent by Alice, the signature verification process will be
terminated, and the signature will be declared invalid. Otherwise,
continue with the steps which follow (ii and iii). Similarly, Alice receives the messages $S_B$, $S_C$ by using the key $K_{AC}$ to decrypt $S_2$. If
$S_B$ in the proxy signature dose not match that sent by Bob,
The signature verification process will be terminated, and the
signature will be declared invalid. Otherwise, continue in accordance with the following steps.

(ii) According to $S_A$ and $S_C$, Bob performs an appropriate
unitary operation on particle 2 to replicate the unknown state
$|\phi\rangle_{A'}$ which carries messages. Similarly, according
to $S_B$ and $S_C$, Alice performs an appropriate unitary operation
on particle 5 to replicate the unknown state
$|\phi\rangle_{B'}$ which carries messages. Alice and Bob
then sends the states $|\phi\rangle_{B'}$ and $|\phi\rangle_{A'}$ to
Trent.

(iii) Trent encodes the quantum state $|\phi\rangle_{A'}$ and
obtains the message $M'''$. He then compares it with $M'$. If
$M'''=M'$, he confirms a series of signatures and messages $(m, S_A,
S_C)$. Otherwise, Trent rejects it. Similarly, Trent encodes the
quantum state $|\phi\rangle_{B'}$, obtains the message
$M''''$ and compares it with $M''$. If $M''''=M''$, he confirms a
series of signatures and messages $(m, S_B, S_C)$. Otherwise, Trent
rejects it.

\section{Security Analysis and Discussion of the Second Scheme}
In this section, we will demonstrate that our scheme meets the following security requirements
\subsection{Message Blindness}
Alice blinded the message $m'$ to $M' $, and Bob blinded the message $m''$ to $M''$. Proxy signer Charlie was unable to obtain the contents of the message because Alice and Bob used
hash functions and XOR operations to blind messages. Therefore, Charlie cannot know the contents of the information he signed.
\subsection{Impossibility of Denial}
We certified that Alice and Bob could not refuse their delegations, nor could Charlie refuse his signature. If the signature is verified, the signer cannot reject its signature or message.
 This is owing to the facts that 1) the signature is encrypted by the key, and 2) the entangled states have stable coherence among particles. Teleportation can be achieved only by measuring
 their own particles. All keys are distributed through the QKD protocol, which has been unconditionally proven. All messages are sent through secure quantum channels. Therefore, Alice and
 Bob cannot deny their authorization, and Charlie cannot refuse his signature.
\subsection{Impossibility of Forgery}
Firstly, we can prove that Charlie cannot forge Alice's or Bob's signatures because the trusted Trent can prevent these attacks in our scheme. Suppose that Charlie tries to forge Alice's
signature. Even if he knows the key, he cannot forge any signature because Charlie's forgery attack will be detected by Trent. Likewise, Charlie cannot forge Bob's signature. In other words,
internal attackers cannot forge their signatures.

Secondly, we assumed that there was an external attacker (Eve). Eve cannot obtain the secret key. Therefore, she cannot forge their signatures. We also assumed that Eve could obtain the secret
key randomly and generate a valid signature with the probability
${\frac{1}{2^n}}$ tending to zero when $n$ tends to infinity. At the same time, we realized the eavesdropping detection function in the construction stage of the quantum channel. Therefore,
 the signing process will only continue when the channel is secure; otherwise, the signing will terminate. Therefore, Eve cannot forge their signatures.

\section{Conclusions}
In this study, we proposed two quantum proxy blind signature schemes
based on controlled quantum teleportation. We presented a scheme for
teleporting an unknown two-particle entangled state with a message
from a sender (Alice) to a receiver (Bob) via a six-particle
entangled channel. Additionally, we presented a scheme for teleporting an unknown, one-particle entangled state with a message passed in both
directions between a sender (Alice) and a receiver (Bob) via a
five-qubit cluster state. Our scheme used a one-way Hash function
and XOR operations to blind the message. The security of the proposed
schemes are guaranteed by the quantum one-time pad and quantum key
distribution, which are different from the previous signature
schemes in classical cryptography. In addition, compared with the existing quantum signature schemes, our schemes have some advantages.
Firstly, unlike the quantum blind signature scheme, both schemes arranged eavesdropping checks to ensure security.
Secondly, the first scheme used quantum teleportation to transmit two unknown qubit states, and were effective.
Thirdly, owing to the maximum connectivity of cluster states and the persistence of entanglement, the second scheme was more secure.
Fourthly, our scheme adopted BM and single-event measurements, which are easy to realize based on the existing technology and experimental conditions.
Therefore, our scheme had a better security profile and could be applied to e-payment, e-commerce, e-voting, and other application scenarios.

However, the following problems still exist in quantum teleportation. 1) In quantum-teleportation experiments,
the first step needs to prepare the quantum channel (which is essentially the preparation of the quantum entangled state);
the technology for preparing the multiparticle entangled state needs to be improved. 2) Moreover, when conducting quantum teleportation,
 the measurement of quantum states is a mandatory step, for which a high-precision measuring instrument is required.
3) Finally, when realizing the teleportation of quantum states, the particles
need to pass through the quantum entanglement channel; during the process of transmission, various interference phenomena will occur. In other words, the teleportation process needs
to be perfected to keep particles from being affected by external interference or physical factors in the process of transmission. The most prominent problem is the security problem
after the combination of quantum teleportation and quantum signature. In the signature, we must prevent outside interferences, the infliction of harm on internal actors, and more importantly,
 the leakage of information. We believe that these problems can be overcome with further investigations.

\bigskip

Acknowledgments: We thank Shao-Ming Fei, Qin Li, and Fei Gao for
helpful discussions. This work was supported by the National
Natural Science Foundation of China under Grant Nos. 12126314,
12126351, and 11861031, and by the Hainan Provincial Natural Science
Foundation of China under Grant No.121RC539. This project was also
supported by the specific research fund of the Innovation Platform
for Academicians of Hainan Province under Grant No. YSPTZX202215.

\bibliography{achemso-demo}

\end{document}